\documentclass[12pt]{iopart}

\usepackage{iopams}

\usepackage{mathrsfs}
\expandafter\let\csname equation*\endcsname\relax
\expandafter\let\csname endequation*\endcsname\relax

\usepackage[T1]{fontenc}
\usepackage{graphicx}
\usepackage{xcolor}

\usepackage[locale=UK,
number-mode=math,
unit-mode=text,
per-mode=symbol,
number-unit-product=\ ,
retain-zero-exponent=false]{siunitx}[=v2]
\DeclareSIUnit{\pixel}{px}
\DeclareSIUnit{\fps}{fps}

\newcommand{\kindex}[2]{\ensuremath{{#1}_{\scalebox{0.65}{#2}}}}
\usepackage{tikz}
\usetikzlibrary{shapes}

\newcommand{\ledge}{\raisebox{0.5pt}{\tikz{\node[scale=0.4,regular polygon, regular polygon sides=60,  fill=black!30, draw=black, line width=0.8pt](){};}}}
\newcommand{\tedge}{\raisebox{0.5pt}{\tikz{\node[scale=0.4,regular polygon, regular polygon sides=4, fill=black, draw=none](){};}}}

\usepackage[noabbrev]{cleveref}

\graphicspath{{../figurematter/latex/figs/}}

\begin{document}

\title[To tread or not to tread]{To tread or not to tread: comparison between water treading and conventional flapping wing kinematics}

\author[Krishna et al.]{Swathi Krishna$^{1,2}$, Alexander Gehrke$^{1}$, \& Karen Mulleners$^{1}$}

% OCID: Karen Mulleners https://orcid.org/0000-0003-4691-8231

\address{$^1$ École polytechnique fédérale de Lausanne (EPFL),  Institute of Mechanical Engineering, 1015 Lausanne, Switzerland}
\address{$^2$ University of Southampton, Department of Aeronautics and Astronautics, Southampton SO16 7QF, United Kingdom}

\ead{s.b.krishna@soton.ac.uk}
 	\begin{indented}
		\item[]July 2022
	\end{indented}

% \date{\textbf{Received:} XX 2020; \textbf{Revised:} XX XX 2020; \textbf{Accepted:} XX XX 2020}

\begin{abstract} % max 200 words
Hovering insects are limited by their physiology and need to rotate their wings at the end of each back and forth motion to keep the wing's leading edge ahead of its trailing edge. The wing rotation at the end of each half-stroke pushes the leading edge vortex away from the wing which leads to a loss in the lift. Unlike biological fliers, human-engineered flapping wing micro air vehicles have different design limitations. They can be designed to avoid the end of stroke wing rotation and use so-called water-treading flapping kinematics. Flapping wings using conventional flapping kinematics have a designated leading and trailing edge. In the water-treading mode, the role of the leading and trailing edges are continuously alternated throughout the stroke. Here, we compare velocity field and force measurements for a rectangular flapping wing conducting normal hovering and water-treading kinematics to study the difference in fluid dynamic performance between the two types of flapping kinematics. We show that for similar power consumption, the water-treading mode produces more lift than the conventional hovering mode and is 50\% more efficient for symmetric pitching kinematics. In the water-treading mode, the leading edge vortex from the previous stroke is not pushed away but is captured and keeps the newly formed leading edge vortex closer to the wing, leading to a more rapid increase of the lift coefficient which is sustained for longer. This makes the water-treading mode a promising alternative for human-engineered flapping wing vehicles.
\end{abstract}

\noindent{\it Keywords}:  Bio-inspired propulsion, Flapping wings, Hovering, Water-treading
%\maketitle

% \ioptwocol % for twocolumn template
\section{Introduction}
\label{sec:intro}

Flapping wing flight continues to inspire engineers to create aerial vehicles with better flight characteristics than steady wing aircraft.
Several multi-disciplinary research activities inspired by bird and insect flight have led to successful demonstrations of flapping wing vehicles.
Some examples at roughly insect scales are the Delfly \cite{DeCroon2012, DeCroon2016}, Robobee \cite{Wood2008,Jafferis2019}, robotic hummingbird \cite{Tu2020}, TL-Flowerfly \cite{Nguyen2015}, four-winged flappers \cite{Ramiro2015} etc.
Specifically, in the insect flight regime, flapping wings are driven by different mechanisms such as a system of gears and servo motors~\cite{DeCroon2012,Hines2014}, torsional springs~\cite{Moore2015}, electromagnetic actuators~\cite{Roll2015, Zou2016}, or piezoelectric actuation~\cite{Ozaki2018}.
Several lab-based experiments predominantly use gear-based mechanisms to achieve flapping kinematics \cite{Dickinson1999, Lua2010, Percin2015,  Roh2017, Krishna2018, Gehrke2021}.
These mechanisms, though inspired by natural fliers, are vastly different in terms of their capabilities and limitations compared to the musculoskeletal configurations of biological fliers.
Such mechanisms provide the possibility of modifying the kinematics of the wing in ways that may not be possible in natural fliers.

The wing kinematics of flying insects during hover are more complex than what most unmanned aerial vehicles are capable of, and can typically be divided into normal hovering and inclined hovering \cite{weis1973, Wang2004a}.
The term normal hovering was coined by Weis-Fogh \cite{weis1973}.
He observed that small hovering insects like fruit flies and bees beat their wings in an almost horizontal plane with symmetric front- and backstrokes.
Normal hovering is used to refer to these flapping kinematics conducted in a horizontal stroke plane that allow small insects to hover and remain at a fixed location in still air.
Some larger insects like dragonflies, employ inclined hovering, where they move their wings back and forth - or up and down - along an inclined stroke plane.
Most birds and bats additionally flap their wings in an asymmetrical way with a long and powerful downstroke during which most of the fluid forces are generated and a short recovering upstroke during which little force is generated \cite{Izraelevitz2014}.

The normal and inclined hover have received considerable attention in the past decades and serve as inspiration for mechanical flapping wing vehicles.
Normal hovering kinematics are most commonly used in lab-based robotic flapping mechanisms~\cite{vandenberg1997, Ansari2009, Lua2015, Krishna2018, Gehrke2018}, whereas inclined hovering has mostly been studied numerically \cite{Wang2004a, Mou2012, Sudhakar2010, cwang2016}.
These classical hovering kinematics always have a rotation of the wing around the pitching axis at the end of a half-stroke to maintain a favourable angle of attack in the subsequent half-stroke.

Aerodynamically speaking, each half-stroke is characterised by the formation of a large-scale leading edge vortex, which is generated at the beginning of the half-stroke and provides a major contribution to the lift generated by the flapping wings \cite{Wang2000, Eldredge2019}.
During the flapping motion, this leading-edge vortex grows along the length of the chord and remains a large-scale feature until the rotation at the end of the half-stroke commences.
This rotation requires additional power and leads to the loss of the leading edge vortex through lift-off, breakdown and decay~\cite{Krishna2018}.
Biological fliers are limited to this kind of rotation at the end of each half stroke due to their skeletal and muscular structures.
They do take advantage of their flexible wings and to maintain positive wing camber during the front and backstroke.
This wing deformation is mostly passive and leads to a delay in stall and enables higher lift coefficients and lower power consumption \cite{yin_effect_2010, eldredge_roles_2010, reid_wing_2019, gehrke_aeroelastic_2022}.

Human-engineered micro air vehicles have different limitations in terms of joints and activation, and instead of exactly mimicking their natural counterparts, they could also be designed to avoid the end of stroke rotation.
Investigation into this idea brings us to the third type of hover that is not observed in natural flight and has received considerably less attention.
This kind of hovering is called the water-treading mode \cite{Freymuth1990, Bai2007a, Tang2008}.
In the normal hover mode, the wings have a designated leading and trailing edge.
In the water-treading mode, the role of the leading and trailing edges are continuously alternated throughout the stroke.
At the end of the half-stroke, the wing is rotated down such that the chord aligns with the stroke plane and the leading edge becomes the trailing edge and vice-versa in consecutive half-strokes.
This modified hover mode eliminates the bluff body dynamics observed in biological hover modes, which means that the leading edge vortex generated in one half-stroke can be moved over to remain atop the suction side for at least part of the next half-stroke.
Many insect wings have a rigid, straight leading edge and a curved, flexible trailing edge.
Due to this chordwise asymmetry, the water-treading mode would not benefit from specialised leading and trailing edges.
However, the design of a passively or actively deforming flexible wing for human-engineered vehicles is challenging.
The water-treading mode would alternatively allow for the use of rigid cambered wings that would be easier to design and fabricate and still provide a positive camber on the front- and the backstroke.

The water-treading hover mode was first proposed by Freymuth who experimentally investigated the qualitative thrust generation for different hovering modes \cite{Freymuth1990}.
In this early experimental study on hover flight, flow visualisation by means of the titanium-tetrachloride was used to compare the flow topology for normal hover and water-treading modes at a Reynolds number of \num{1700}.
Freymuth revealed the existence of dynamic stall vortices for thrust generation and  extraordinarily high thrust coefficients in the range from \numrange{5}{7} were found in both hover modes \cite{Freymuth1990}.
No information on the efficiency was found in this work.

The study of the water-treading mode seemed to remain dormant after Freymuth's initial study \cite{Freymuth1990} and resurfaced close to two decades later \cite{Berman2007, Bai2007a, Tang2008}.
The aerodynamic performance of the water-treading mode was compared to the performance of the normal hovering kinematic mode based on numerical simulations for an elliptical airfoil for different Reynolds numbers and reduced frequencies \cite{Tang2008}.
The water-treading mode was shown to yield higher lift and lower drag compared to the normal hover kinematics.
Switching leading and trailing edges between each stroke opens up the possibility to use cambered stiff wings to further increase the aerodynamic performance of flapping wings \cite{Isaac2008}.
Further, the water-treading mode (referred to as the bionic mechanism in this study) was compared with the flapping motion of a fruit fly \cite{Bai2007a}.
The ratio of the mean lift to the mean drag of the water-treading motion was \SI{35.0}{\percent} greater in the advanced pitch, \SI{66.1}{\percent} greater in the symmetrical pitch, and \SI{150.0}{\percent} greater in the delayed pitch when compared to the performance of the fruit fly hovering kinematics.
More recently, detailed flow features from numerical simulations on a three-dimensional wing were correlated with the lift and power requirements for a wide range of parameter variations, confirming the advantages of the water-treading motion~\cite{Lua2017}.
In this study, it was suggested that natural fliers rely on normal hover despite the advantages of the water-treading mode because it may require less power to pitch the insect wings down at the start of the stroke than to pitch them up.
This was also suggested by Berman and Wang \cite{Berman2007} who used a combination of a genetic algorithm and a gradient-based optimisation to find energy minimising flapping kinematics for hovering flight.
A common feature among the energy-minimising kinematics they found was their tendency to maintain the leading edge throughout the flapping cycle and not alternate the leading and trailing edges as it is the case in the water-treading mode.
However, the authors used a quasi-two-dimensional model to predict the forces on the hovering wings, which does not account for the presence of vortices.
In one of the latest studies, honeybees were found to adopt a motion similar to water-treading on the surface of water \cite{Roh2019}.
Even though these studies have presented a wealth of data by varying the kinematic and morphological parameters, the exact driving mechanisms that can provide the aerodynamic advantage of the water-treading mode deserve further attention.
% Therefore, these promising kinematics deserve to be revisited especially in view of their application potential.

In this study, we performed an experimental comparison of the flow and force evolution for normal hovering and water-treading kinematics.
We especially focus on the differences in the vortex formation and wing-vortex interaction at stroke reversal for both types of kinematics.
We hypothesise that the aerodynamic power required for the water-treading mode will be lower than for the normal hovering mode because the conventional rotation against the direction of stroke motion at stroke reversal is eliminated.
Furthermore, we expect higher lift values for the water-treading mode than in the normal hovering mode as the water-treading kinematics will retain the leading edge vortex closer to the wing and inhibits its full breakdown and decay.
These hypotheses will be tested using time-resolved force and moment measurements combined with phase-locked particle image velocimetry.

\section{Experimental set-up}
\label{sec:exp}

\subsection{Experimental hardware}
\begin{figure}
\centering
\includegraphics{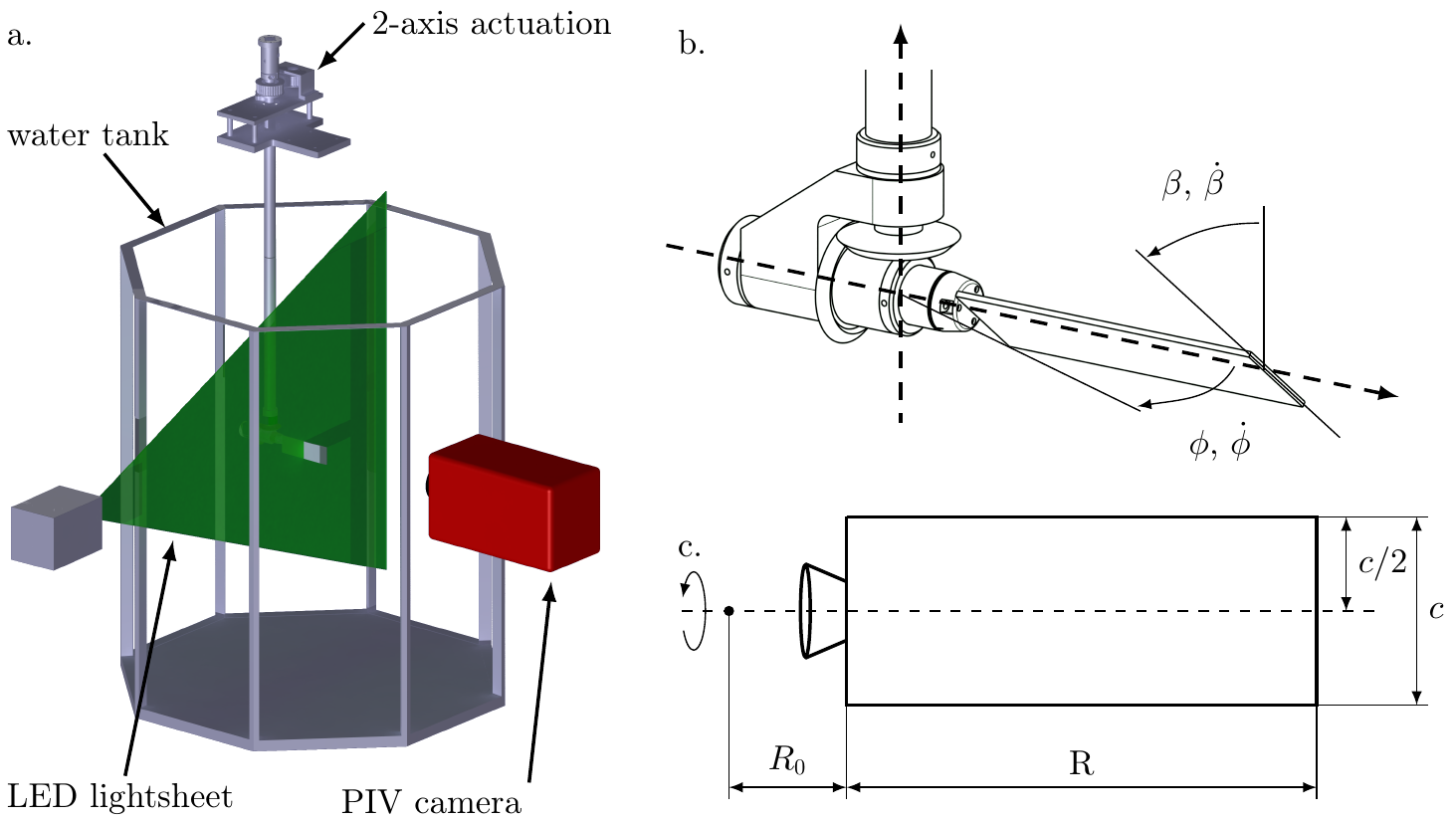}
\caption{
	(a) Overview of the experimental configuration.
	(b) Stroke $\phi$ and pitch angles $\beta$ defining the flapping wing kinematics, and
	(c) Rectangular wing dimensions and pitch axis location.}
\label{fig:expSetup}
\end{figure}

Phase-locked particle image velocity (PIV) and direct force measurements were carried out on a mechanical flapping wing model in a quiescent flow (\cref{fig:expSetup}a).
The experimental set-up and procedure are the same as described in \cite{Gehrke2021} and will be summarised here.

A rigid rectangular planform is adopted for the wing with a span $R=\SI{107}{\milli\meter}$, chord $c=\SI{34}{\milli\meter}$, and a thickness-to-chord ratio of \num{0.05} shown in \cref{fig:expSetup}c.
The wing is hinged at the mid-chord position to a six-axis IP68 force-torque transducer (Nano17, ATI Industrial Automation, USA), which is connected to the flapping wing mechanics that control the stroke and pitch motion.
Stroke refers to the reciprocating motion of the wing and wing pitch refers to the change in angle of attack (\cref{fig:expSetup}b).
The stroke and pitch motion are driven by two servo motors (Maxon motors, type RE35, \SI{90}{\watt}, \SI{100}{\newton\milli\meter} torque, Switzerland) reduced by $35:1$ with a planetary gear-head for the stroke and $19:1$ for the pitch actuation.
The mechanism is placed inside an octagonal tank with a diameter \SI{750}{\milli\meter} filled with a mixture containing a volume percentage of \SI{65.0}{\percent} glycerine and \SI{35.0}{\percent} water.
At a temperature of $\SI{21}{\celsius}$, the glycerine-water mixture has a density of $\rho = \SI{1180.4}{\kilogram\per\meter\cubed}$ and a kinematic viscosity of $\SI{18.77e-6}{\metre\squared\per\second}$.

The two non-dimensional parameters associated with flapping flight in hover are the reduced frequency ($ k $) and the Reynolds number ($ Re $).
The reduced frequency for the model wing is given by $k= \pi c/ 2 \phi \kindex{R}{2}  $, where $2 \phi$ is the peak-to-peak stroke amplitude, $\kindex{R}{2}= \sqrt{1/R\int_{\kindex{R}{0}}^{\kindex{R}{0}+R}r^2 dr} $
is the radius of the second moment of area, and $ \kindex{R}{0} $ is the distance between the stroke axis and the wing root indicated in \cref{fig:expSetup}c.
The Reynolds number is defined as $ Re= \bar U c/ \nu $ , where $ \nu  $ is the kinematic viscosity of the fluid, $ \bar U =2 \phi f \kindex{R}{2}$ is the characteristic velocity and $f$ is the flapping frequency.
For $f=\SI{0.25}{\hertz}$, we obtain a reduced frequency $ k=\num{0.37} $ and a Reynolds number of $Re=\num{130}$ for the given configuration.

Phase-locked measurements are performed following the same procedure as adopted in our previous work \cite{Krishna2018}.
A total of \num{30} stroke cycles were carried out.
The data recorded from the first \num{5} cycles were eliminated to remove transient effects from the start-up of the motion.
A cylindrical lens is used to create a light sheet of approximately \SI{4}{\milli\meter} thickness from pulsed light emitting diodes (LED) that operate at a wavelength around \SI{530}{\nano\meter} (LED Pulsed System, ILA\_5150 GmbH, Germany).
The two LED light sheets are carefully aligned to illuminate fluorescent dye particles in the flow field from opposite directions.
The illuminated plane is recorded by a sCMOS camera with a \SI{2560x2060}{\pixel} resolution (ILA\_5150 GmbH / PCO AG, Germany) covering a \SI{109x94}{\milli\meter} field of view.
The raw data are processed with a multi-grid algorithm with a resulting interrogation window size of \SI{32x32}{px} and an overlap of \SI{50}{\percent} is used to correlate the raw images.
This yields a physical grid resolution of \SI{1}{\milli\meter} or \SI{0.034}{c} in the resulting velocity fields.

The forces are recorded via a data acquisition card (National Instruments, USA) with a sampling frequency of \SI{1000}{\hertz}.
The force data in the time-resolved plots were filtered with a zero phase delay low-pass $5^{th}$ order digital Butterworth filter.
The cut-off frequency was chosen to be \num{12} times the flapping frequency.
The load cell is located at the wing root and pitches along with the wing.
The force and power coefficients of the system are calculated from the force and torque measurements according to:
\begin{equation}
	\kindex{C}{L}= \frac{L}{0.5 \rho c R \bar U^2}, \hspace{10 mm} \kindex{C}{P}=\frac{P}{0.5 \rho c R \bar U^3},
\end{equation}
where $ L $ is the instantaneous lift and $ P $ is the aerodynamic power of the system.
For the two-axis motion, total power is defined as the sum of the pitching power and the stroke power and is calculated as described in \cite{Gehrke2021}.
The hovering efficiency of the flapping wing ($ \eta $) is calculated as the ratio between the stroke average lift and power coefficient:
\begin{equation}
	\eta = \frac{\bar{\kindex{C}{L}}}{\bar{\kindex{C}{P}}}\quad.
\end{equation}

\subsection{Hovering kinematics}
\label{sec:kine}

\begin{figure*}
\centering
\includegraphics{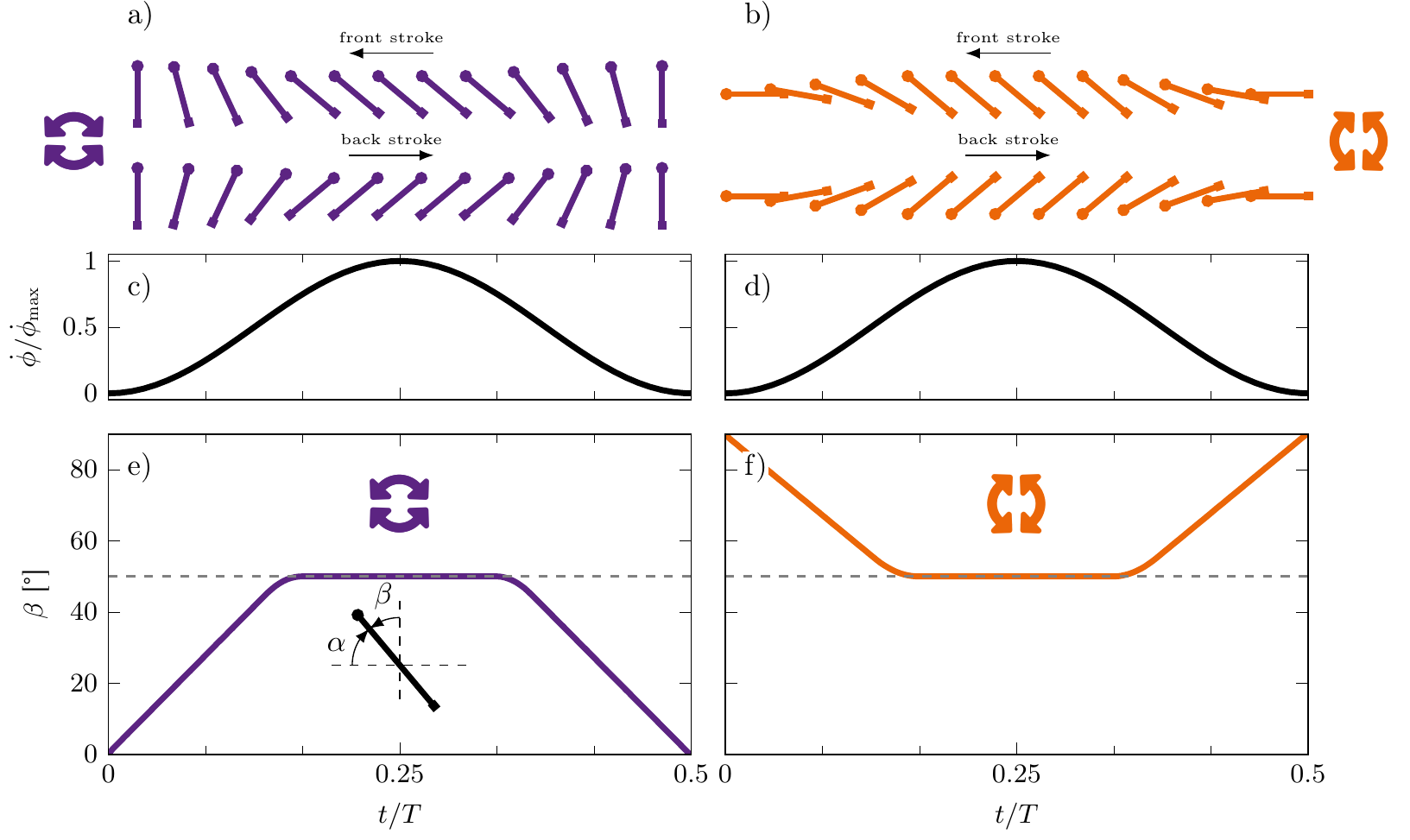}
\caption{Conceptual representation of the symmetric normal hover and water treading wing kinematics (a,b).
Temporal evolution of the stroke velocity (c,d), and pitch angle (e,f).
The duration of wing pitch is $\kindex{T}{f}=T/3$ with $T=4 s$.}
\label{fig:kine1}
\end{figure*}

For the direct comparison between the normal hovering and the water-treading kinematics, a standard symmetric pitching motion is considered for both hovering modes.
Conceptual representations of the two symmetric insect-inspired and water-treading kinematics are represented in the top row of \cref{fig:kine1}.
In the classical hovering scenario, there is a designated leading edge (\ledge) which consistently proceeds the designated trailing edge (\tedge) during the front and the backstroke.
In the water-treading mode, the edge of the wing that leads during the front stroke (\ledge), lags behind the other edge (\tedge) during the backstroke and vice-versa (\cref{fig:kine1}).
We say that the leading and trailing edge alternate roles between half-strokes.

The schematics of the stroke and pitching motions are represented in \cref{fig:kine1}.
In the stroke plane, the wing moves in a sinusoidal motion for both hovering modes.
The angle of attack ($ \alpha $) is defined as the angle over which the leading edge must be pitched in the direction of the stroke motion such that the leading edge would align with the stroke plane.
The angle of attack is the same for both modes during the translation phase. %where most of the lift and power is being generated.
The stroke amplitude and pitch amplitude values for the base case are chosen to mimic that of a hoverfly \cite{Walker2010, Mou2011}, similar to our previous work \cite{Krishna2018, Krishna2019}.
The duration of a single pitch manoeuvre in the current study is $ \kindex{T}{f}=T/3 $, where $T=\SI{4}{\second}$ is the period of a flapping cycle.

\section{Results and analysis}
\label{sec:res}

\subsection{Power and lift}
\label{sec:sym}

\begin{figure*}
\centering
\includegraphics{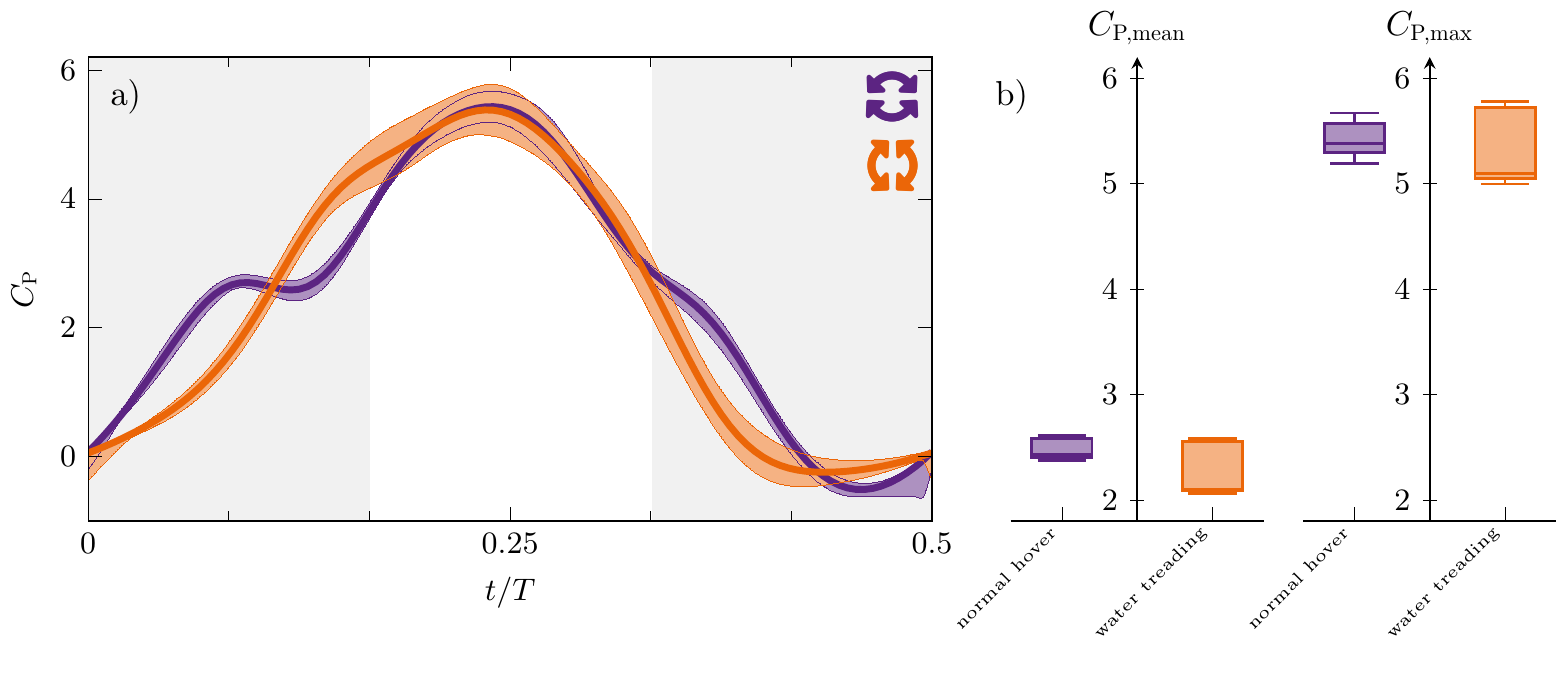}
\caption{Temporal evolution of the phase averaged power coefficient ($ \kindex{C}{P} $) in a half-stroke for symmetric normal hover and water-treading motions (a).
The grey regions indicate the duration of the rotation.
Direct comparison of the overall mean and maximum power coefficient for both hovering modes in terms of box plots (b).
The whiskers extend from the minimum to the maximum values observed across the ensemble of measured flapping strokes.
The box itself covers the interquartile range of the measured values.
The horizontal line across the box marks the median value.
}
\label{fig:cpsym}
\end{figure*}

To analyse the performance of the two hovering modes, we first compare the temporal evolution of their power and lift coefficients for a symmetric flapping motion as presented in \cref{fig:kine1}.
Due to the symmetry of the prescribed motion and thanks to the high precision of our experimental flapping wing set-up, we obtain quasi-identical force and torque responses during the front- and the backstroke.
The phase averaged aerodynamic loads are presented for one-half cycle, which corresponds to a single back- or front-stroke.
They are obtained by averaging over all \num{50} half-strokes from \num{25} flapping cycles.
The phase-averaged curves are surrounded by a shaded region that indicates the envelope bounded by the maximum and minimum values across the ensemble of recorded half-strokes.

The temporal evolution of the coefficient of power ($\kindex{C}{P}$) for the normal hovering and water-treading modes is presented in \cref{fig:cpsym}.
The grey area indicates the duration of the rotation.
There are two regions where the wing rotates, one at the beginning and one at the end of the half-stroke.

The power coefficients for both normal hover and water-treading modes start and end at zero, and reach approximately the same maximum value of $\kindex{C}{p}\approx\num{5.4}$ at the middle of the half stroke around $ t/T=0.24$.
The normal hover requires only marginally higher maximum power ($\approx$ \SI{1}{\percent}) than the water-treading mode, which is indicated by the box plots in \cref{fig:cpsym}b.
The stroke averaged power is about \SI{7}{\percent} higher for the normal hover than the water-treading mode but remains within the interquartile range of the water-treading mode results.
The biggest difference between the two modes occurs during the initial rotation of the wing (\cref{fig:cpsym}a).
In the water-treading mode, power increases slowly but continuously during the initial rotation and remains below the power curve of the normal hovering.
The power required during normal hover is higher than that of the water-treading mode at the very beginning of the stroke but ceases to increase around $t/T \approx 0.08$.
The power coefficient remains around $\kindex{C}{p}\approx\num{0.24}$ until the end of the initial rotation ($t/T \approx 0.16$) and it increases again thereafter.
Around $t/T \approx 0.13$, the power curves cross, and the water-treading mode requires slightly more power than the normal hover mode during the first part of the translation phase of the cycle.
The translation phase of the cycle is the portion of the cycle where the wing has a constant angle of attack and the power is mainly influenced by the stroke velocity which is the same for both motions.
The power curves for both motions overlap for $0.2 \leq t/T \leq 0.33$, which lies within the translation phase, and they reach a similar maximum value around \num{5.4} (\cref{fig:cpsym}a).
The normal hovering mode requires slightly more power than the water-treading mode during the end of stroke rotation.

The differences in the mean and maximum stroke averaged power between the two modes are marginal and of the same order of magnitude as the differences measured between flapping strokes.
Contrary to our initial expectation, the overall power requirement in the water-treading mode is not substantially lower than that of the normal hover.
Subtle differences are observed in the temporal evolution of the power during the rotation at the start of the stroke.

\begin{figure*}
\centering
\includegraphics{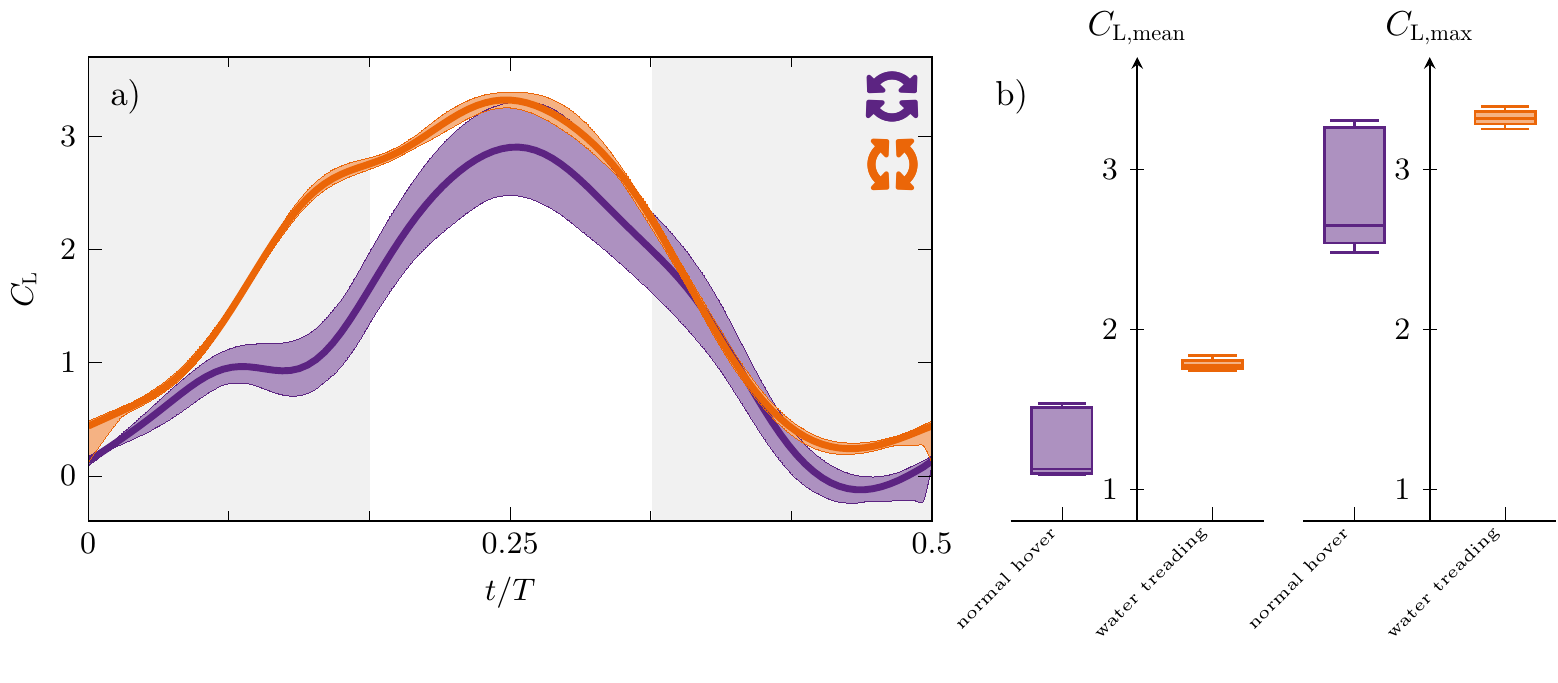}
\caption{Temporal evolution of the phase averaged lift coefficient ($ \kindex{C}{L} $) in a half-stroke for symmetric normal hover and water-treading motions (a).
Direct comparison of the overall mean and maximum lift coefficient for both hovering modes in terms of box plots (b).
The whiskers extend from the minimum to the maximum values observed across the ensemble of measured flapping strokes.
The box itself covers the interquartile range of the measured values.
The horizontal line across the box marks the median value.
}
\label{fig:clsym}
\end{figure*}

The largest differences in the temporal lift responses are also observed during the rotation phase at the start of the stroke for $0.08 \lesssim t/T \lesssim 0.18$ (\cref{fig:clsym}a).
From the start of the stroke, the water-treading mode yields a higher lift compared to the normal hovering mode.
The lift generated during normal hover initially also increases in time at a comparable rate as for the water-treading mode, but it stagnates between $t/T \approx 0.08$ and the end of the initial rotation.
The difference between the two modes is largest during this time period.
Once the pure translation begins, the lift increases faster for the normal hover than for the water-treading mode.
In both cases, a maximum in the lift coefficient is reached at $t/T=0.25$, and the lift drops in a similar way in both hover modes in the second part of the half-stroke.
The water-treading mode yields a \SI{12}{\percent} higher stroke average maximum lift than the normal hover.
The higher stroke-to-stroke variations during normal hover diminishes the significance of this gain.
We do measure a significant increase in the stroke average mean lift of \SI{25}{\percent} for the water-treading mode with respect to the normal hovering mode.
The water-treading mode is thus more advantageous than the normal hover in terms of overall lift generation.
This lift increase comes at no additional cost in terms of power requirement.

\begin{figure*}
\centering
\includegraphics{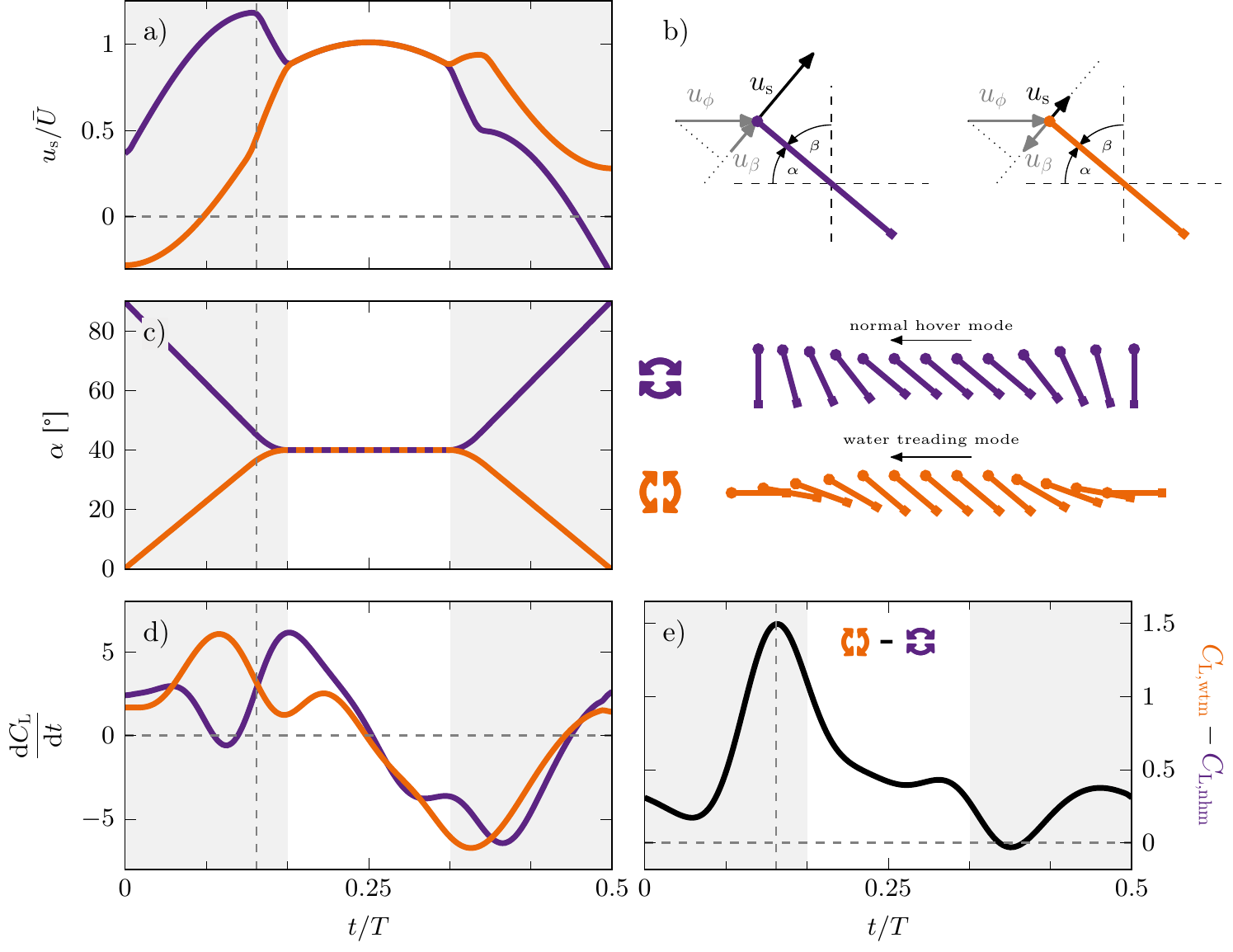}
\caption{(a) Evolution of the shear layer velocity ($ \kindex{u}{s} $) normalised with average stroke velocity ($\bar{U}$), (b) schematic representing the definition of the shear layer velocity, (c) temporal evolution of the angle-of-attack, (d) temporal evolution of the rate of change of lift coefficient in both hover modes, and (e) the difference between the lift coefficient evolutions of both hover modes.
The dashed vertical line marks the timing of the maximum difference in lift between the hover modes and coincides with the maximum shear layer velocity during normal hover.}
\label{fig:moreloads}
\end{figure*}

To understand the origin for the enhanced lift generation in the water-treading mode, we first investigate the effect of kinematics on the lift production.
In recent work on the optimisation of pitching kinematics of a flapping wing, Gehrke and Mulleners \cite{Gehrke2021} demonstrated that the leading edge shear layer velocity serves as the characteristic scalar quantity that governs the force response of arbitrary pitching motions.
The leading edge shear layer velocity is directly extracted from the input kinematics and computed as the chord-normal projection of the velocity of the leading-edge due to the stroke and pitch motions at the span-wise location \kindex{R}{2}, which corresponds to the second moment of area of the wing, such that:
\begin{equation}\label{eq:us}
	\kindex{u}{s}(t)=\kindex{R}{2} \dot \phi (t) \cos \left(\beta(t)\right) +0.5 c \dot \beta(t)\quad.
\end{equation}

The temporal evolution of the shear layer velocity, normalised by the average stroke velocity, $\bar{U}$, is presented in \cref{fig:moreloads}a.
A schematic highlighting the two different components that contribute to \kindex{u}{s} is shown in \cref{fig:moreloads}b.
At the start of the cycle, the shear layer velocity is dominated by the leading edge pitch rotation and is positive for the normal hovering mode, due to the increase in $\beta$, and negative for the water-treading mode, due to the decrease in $\beta$.
The shear layer velocity for the normal hovering increases sub-linearly until it reaches a maximum value close to the end of the initial rotation at $t/T\approx 0.15$ when the angle of attack has almost reached its target value of \ang{40} (\cref{fig:moreloads}c).
This sub-linear increase in the shear layer velocity has an adverse effect on the lift increase (\cref{fig:moreloads}d).
The lift more or less stagnates for $0.08<t/T< 0.12$ and increases more rapidly at the end of the initial rotation.
In the water-treading case, the shear layer velocity increases super-linearly during the initial rotation which leads to a stronger increase in the lift coefficient at the beginning of the stroke (\cref{fig:moreloads}d).
The maximum rate of change of the lift coefficient is the same for both hover modes but occurs earlier for the water-treading mode than for the normal hovering mode.
This earlier increase in lift gives the water-treading mode the upper hand in terms of lift.
The maximum difference in lift between the hover modes occurs at $t/T\approx 0.15$ and coincides with the time at which the maximum shear layer velocity is reached during normal hover (\cref{fig:moreloads}e).
The higher values of the shear layer velocity in the normal hover do not lead to higher lift, which seems to be more affected by the rate of increase of shear layer velocity.

\begin{figure*}
\centering
\includegraphics{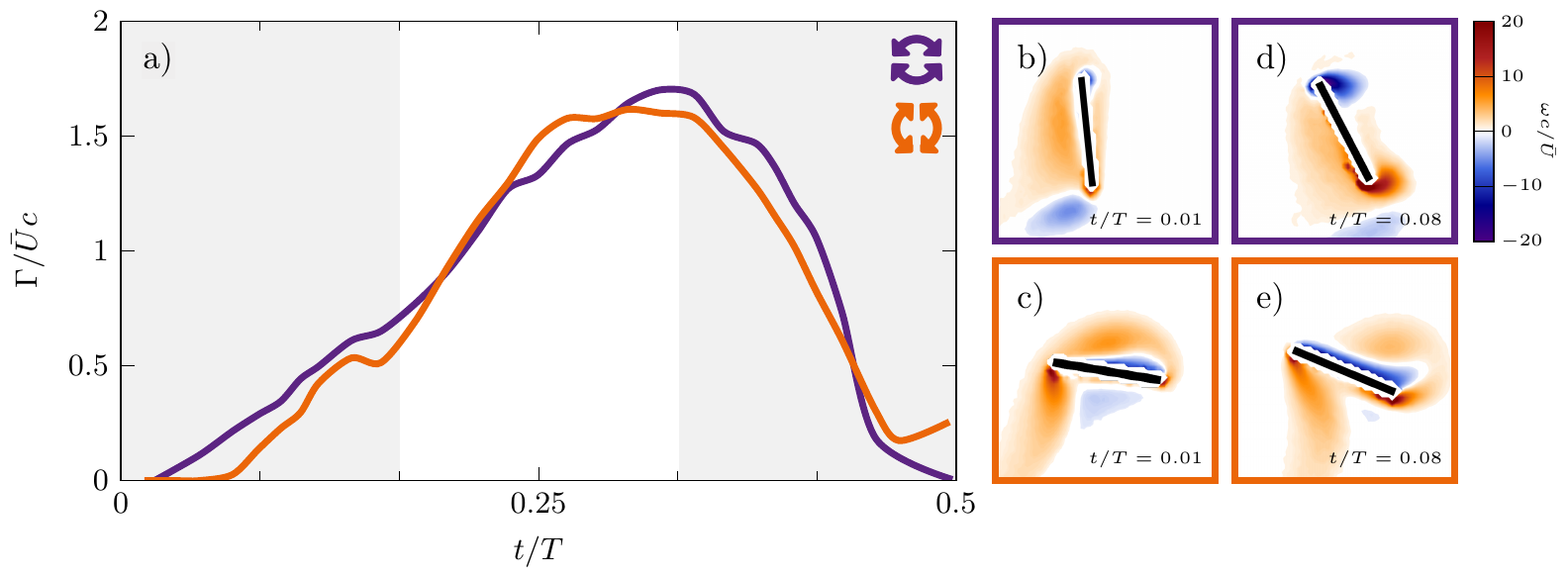}
\caption{(a) Temporal evolution of the leading edge vortex circulation for normal hover and water treading modes, and selected vorticity snapshots at the start of the stroke and the pitch rotations; (b,d) normal hover mode,  (c,e) water-treading mode.}
\label{fig:circvort}
\end{figure*}

To further highlight the dynamic differences between both hover modes, we present the temporal evolution of the leading edge vortex circulation and the vorticity fields at the start of the stroke in \cref{fig:circvort}.
The leading edge vortex circulation is extracted from phase-averaged PIV snapshots and non-dimensionalised by the chord length and the stroke average velocity.
In both cases, we see a similar increase in the leading edge vortex lift and both reach approximately the same maximum value of $\Gamma/(c\bar{U})=1.6$ at the end of the translation phase.
This suggests that the leading edge vortices in both modes reach similar vortex strength before breaking down during the end of stroke rotation, despite the differences in pitch kinematics.
This explains why we do not observe a significant difference in the maximum lift coefficient.
The only minor difference we can note is that the leading edge vortex circulation increases rapidly in the water treading mode after a short delay at the start of the stroke ($0<t/T<0.08$).
This delay is not observed in the normal hover where the circulation increases steadily right from the beginning.
For the former, the angle of attack is zero at the start of the stroke and vorticity is not generated until the angle of attack increases.
The spatial growth of the leading edge vortex at the start of the stroke in the water-treading mode is also limited by the presence of the leading edge vortex from the previous stroke (\cref{fig:circvort}(c)).

To further corroborate this, we use the vorticity fields obtained from PIV in  \cref{fig:circvort}(b)-(e).
The positive (clockwise) vorticity is in blue and the negative (anticlockwise) vorticity is in orange.
The top row represents the normal hover and the bottom row represents the water-treading mode.
The time instants at which the images are shown correspond to the very beginning of the stroke, where the biggest difference between the two modes is observed.

The significant difference between the two modes is in the vorticity distribution on the pressure and suction side of the wing, which arises from the difference in the angle of attack in both modes.
There is little new vorticity at the leading edge on the suction side in the normal and water-treading mode at the start of the half-stroke  at $t/T=0.01$ (\cref{fig:circvort}(b,c)).
Anti-clockwise vorticity from the previous stroke, which was a part of the leading edge vortex in that stroke, surrounds the wing in different ways in both modes at the beginning of a new cycle.
In normal hover, this vorticity sits on the pressure side of the wing (\cref{fig:circvort}b), whereas
in the water-treading mode, the leading edge vortex from the previous stroke cycle is captured and retained on the suction side in the consecutive cycle (\cref{fig:circvort}c).
The distribution of the remnant vorticity from the previous stroke affects the formation of the leading edge vortex in the new half-stroke.

A clear, compact, leading edge vortex emerges in the normal hover mode at $t/T=0.08 $ as expected (\cref{fig:circvort}d).
At the same time instant in the water-treading mode (\cref{fig:circvort}e), there is an absence of a discernible leading edge vortex.
Here, a thin but strong shear layer develops between the suction side of the wing and the captured leading edge vortex from the previous half-stroke  (\cref{fig:circvort}e).
The captured counterclockwise rotating leading edge vortex now develops as the trailing edge vortex in the new stroke and imparts a downward force on the newly developing shear layer, binding it to the wing.
The action of the captured vortex indirectly contributes to the enhanced lift in the water-treading mode along with the new shear layer and form together an enlarged region of low pressure on the wing.
The captured vortex then moves away and the new leading edge vortex grows over the entire chord length, further sustaining the advantage in the lift.

\subsection{Efficiency}
\label{sec:eff}

\begin{figure}
\centering
\includegraphics{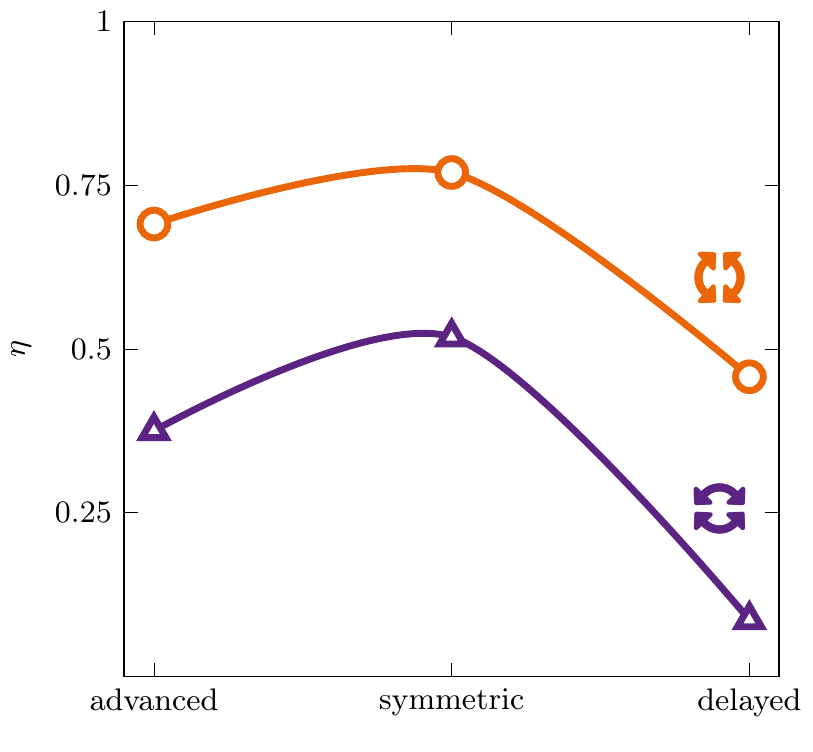}
\caption{Efficiency ($\eta$) of normal hover and water-treading modes for different phase-shifts between the pitch and the stroke motion at a pitch amplitude of $\beta=\ang{50}$.
Advanced means that the pitch motion leads the stroke motion, delayed means that the pitch motion lags the stroke motion.}
\label{fig:eff}
\end{figure}

The water-treading mode generates more lift than the normal hover mode but the average power is nearly the same for both modes.
The efficiency, which is the ratio of the mean lift to mean power, is thus greater for the water treading mode for a flapping wing with symmetric pitch.
The measured values of the fluid dynamic efficiency are presented in \cref{fig:eff} for the two hover modes, a pitch amplitude $\hat \beta=\ang{50}$, and different phase shifts between the pitch and the stroke motion.
These phase-shifts are classified as advanced, symmetric, and delayed pitch \cite{Krishna2019}.
In the symmetric case, half of the wing rotation is executed at the end of the half-stroke and the other half of the rotation is completed at the beginning of the next half-stroke.
In the advanced pitch case, the wing starts to rotate earlier in the half-stroke such that most or even all of the wing rotation is completed prior to stroke reversal.
In the delayed pitch case, the wing starts to rotate later in the half-stroke such that most or even all of the wing rotation is completed after stroke reversal.
Here, we present results for a pitch advancement and delay by $T/6$ for a duration of the wing rotation of~$\kindex{T}{f}=T/3$.
This means that the rotation is fully advanced or fully delayed.

% \begin{figure}
% \centering
% \includegraphics[scale=0.5]{../figurematter/plots/mclmcp}
% \caption{Mean lift and power for advanced, symmetric, delayed rotations to explain the efficiency trends. (Do we need this? )}
% \label{fig:eff2}
% \end{figure}

For the two types of hover kinematics, the symmetric motion is most efficient, followed by the most advanced.
The delayed rotation is least efficient.
Interestingly, the mechanisms behind the increased efficiency seem different in both cases.
For normal hover, the increase in efficiency is mainly driven by a reduction in the mean power.
Both the advanced and delayed rotation require at least \SI{40}{\percent} more power than the symmetric rotation.
The mean lift for normal hover is highest for the most advanced rotation, followed closely by the symmetric rotation.
For the water treating mode, the increase in efficiency is mainly driven by an increase in the mean lift.
Here, the power requirements are highest for the symmetric rotation, but the symmetric rotation also generates higher mean lift compared to both the advanced and the delayed rotation.

Overall, the water-treading mode is significantly more efficient than the normal hover mode for the tested pitch angle $\hat{\beta}$ (\cref{fig:eff}).
For the symmetric rotation and $\hat{\beta}=\ang{50}$, the water-treading mode is about \SI{50}{\percent} more efficient than the normal hover.
The general increase in efficiency for the water-treading mode agrees with Lua et al. \cite{Lua2017}.
Slightly larger gains in efficiency are seen for the fully advanced and the fully delayed rotations (\cref{fig:eff}).

\begin{figure}
\centering
\includegraphics{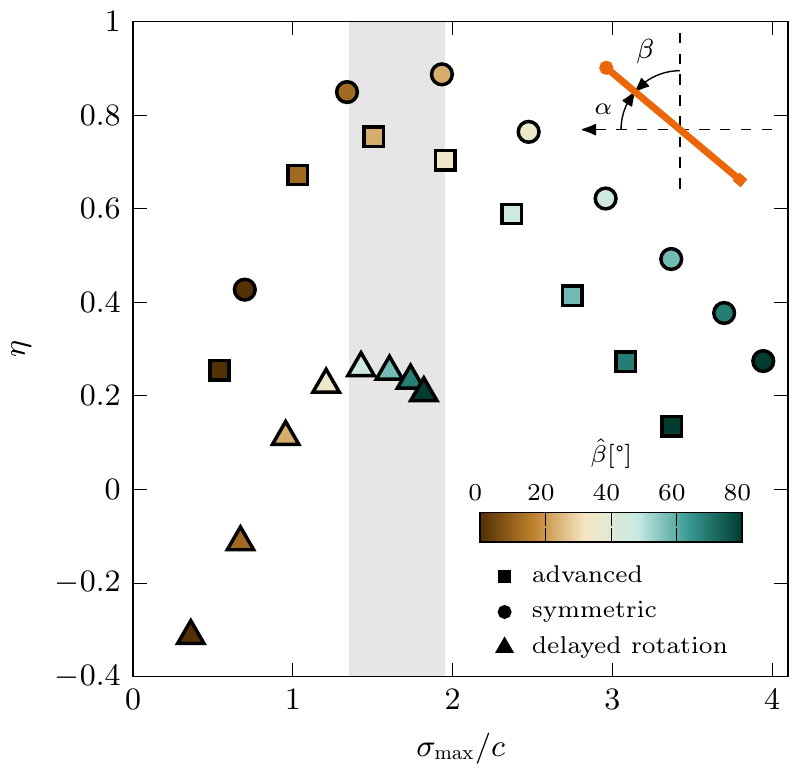}
\caption{Efficiency $\eta$ of the water-treading mode over normalised maximum advective time $\kindex{\sigma}{max}$ for different phase-shifts represented by the symbols. The colour-coding represents the variations in pitch amplitude.}
\label{fig:effvar}
\end{figure}

To identify the optimum efficiency kinematics using the water-treading mode, we varied the pitch amplitude $\hat{\beta}$ ranging from \SIrange{10}{80}{\degree}.
The efficiency for various pitch amplitudes is presented as a function of the maximum advective time in \cref{fig:effvar} for symmetric, fully advanced, and fully delayed pitch motions in the water-treading mode.
% The advective time is a measure of the age of the leading edge vortex.% and the maximum value indicates the maximum non-dimensional time during which vorticity advected by the shear layer.
Previous work by Gehrke and Mulleners \cite{Gehrke2021} has revealed that the advective time serves as the characteristic time scale for the growth of the leading edge vortex and the evolution of the the aerodynamic force.
It is calculated by integrating the shear layer velocity ($\kindex{u}{s}$) from the start of the stroke motion following
\begin{equation}
\sigma (t)=\int\limits_{0}^{t} \kindex{u}{s}(\tau) d\tau \quad.
\end{equation}
The maximum advective time is a measure of how much time the leading edge vortex has to grow during the stroke.

The maximum advective time is influenced by the pitch amplitude and the pitch delay (\cref{fig:effvar}).
Overall, smaller pitch amplitudes lead to larger angles of attack which lead to higher shear layer velocities, faster feeding of the leading edge vortex, and higher values of the maximum advective time.
Delayed rotations delay the start of the formation of the leading edge and reach lower values of the maximum advective time.
Advanced rotations also lead to shorter maximum advective times compared to the symmetric rotations because the advancement of the rotation reduces the angle of attack and the shear layer velocity earlier compared to the symmetric rotation.
The highest values of the maximum advective time are found here for the symmetric rotation.

Overall, the symmetric pitch is the most efficient of the three phase shifts, followed by the advanced and then the delayed pitch.
The efficiency for the delayed rotation cases starts below zero at low values of $\hat{\beta}$ due to the negative mean lift generated here.
For all three phase-shifts, the efficiency initially increases as the pitch amplitude increases and reaches a maximum value at different pitch amplitudes but at similar maximum advective time scales of $\kindex{\sigma}{max}/c$ ranging from \numrange{1.4}{1.9}.
The pitch amplitude that delivers maximum efficiency in the advanced and symmetric case is $\hat{\beta} \approx \ang{30}$.
In the delayed rotation case, a larger pitch amplitude of $\hat{\beta} \approx \ang{50}$ is required to reach peak efficiency.
Further increasing the pitch amplitude leads to larger values of the maximum advective time but a drop in efficiency.

\section{Summary and conclusions}
An alternative type of pitching kinematics for flapping wings in hover, called the water-treading mode, was implemented in this experimental study and its performance was compared with the performance of conventional bio-inspired hover kinematics.
% For a direct comparison, a standard symmetric pitch and fixed stroke motion are selected chosen as the base case.
A combination of velocity field measurements using PIV and direct force measurements were carried out to reveal the physical mechanisms that lead to a change in the aerodynamic performance in the two modes.
The measurements were carried out for a Reynolds number of $Re=130$ and a reduced frequency of $k=0.37$.
Overall, this study delves deeper into the benefits of the water-treading mode in terms of lift production and efficiency.

The stroke averaged power is about \SI{7}{\percent} higher for the normal hover than the water-treading mode for a pitch amplitude $
\hat{\beta}=\ang{50}$ but remains within the interquartile range of the water-treading mode.
The maximum power difference between the two modes is around \SI{1}{\percent}.
The absence of a significant variation in the average and the maximum power coefficients between the different pitching kinematics was surprising and suggests that the power requirement is primarily a function of the stroke velocity.
The difference in the pitching mode only leads to subtle differences observed in the temporal evolution of the power coefficient at the beginning of the stroke.

The major difference between the two pitching modes is observed in the lift coefficient.
The water-treading mode yields a \SI{12}{\percent} higher stroke maximum lift but this is less significant if the stroke-to-stroke variations are considered.
However, a significant \SI{25}{\percent} increase in mean lift is obtained with the water-treading mode.
This is noteworthy as the increase in lift comes at a similar costs of power and leads to a significant increase in efficiency.
Overall, the water-treading mode is more efficient for hovering flight than the normal mode also if the pitching motion is advanced or delayed with respect to the stroke motion.
The efficiency is the highest when the wing pitches symmetrically about the stroke reversal.
For all three phase-shifts considered, the efficiency reaches a maximum value at intermediate but different pitch amplitudes at similar maximum advective time scales of $\kindex{\sigma}{max}/c$ ranging from \numrange{1.4}{1.9}.
The pitch amplitude that delivers maximum efficiency in the advanced and symmetric case is $\hat{\beta} \approx \ang{30}$.
In the delayed rotation case, a larger pitch amplitude of $\hat{\beta} \approx \ang{50}$ is required to reach peak efficiency.

The main contribution of the enhanced lift generated in the water-treading mode is associated with an increase of the shear layer velocity.
The super-linear increase in the shear layer velocity of the water-treading mode leads to a steeper increase in the lift coefficient.
This is due to favourable angle-of-attack as the stroke velocity increases.
Even though the maximum value of the shear layer velocity is higher in normal hover, the continuous  increase in shear layer velocity leads to more sustained build-up of the lift in the water-treading mode.

Snapshots of the flow fields further highlight the benefits of the water-treading motion.
The leading edge circulation is nearly the same for both hover modes, indicating that the leading edge vortices reach similar vortex strength before breaking down during the end of stroke rotation.
This explains the lack of major difference in the maximum lift coefficient.
In the water-treading mode, the favourable angle-of-attack helps the wing to capture the leading edge vortex from the previous half-stroke that serves to keep the newly developed shear layer bound to the wing, leading to an earlier increase of the lift coefficient which is sustained for longer.

Our results indicate that the water-treading mode is a promising alternative flapping wing kinematic mode during hover for the design of human-engineered flapping wing vehicles.

\paragraph{Funding Statement}
This work was supported by the Swiss National Science Foundation under grant number 200021\_175792.

\paragraph{Declaration of Interests}

The authors declare no conflict of interest.

\paragraph{Author Contributions}

S.K. and K.M. created the research plan, designed experiments, and formulated the problem.
S.K. and A.G performed the experiments and analysed the data.
S.K. wrote the manuscript.
A.G. and K.M. edited the manuscript.
K.M. acquired the funding for the project.

\paragraph{Data Availability Statement}
Raw data are available from K.M.

\paragraph{Ethical Standards}
The research meets all ethical guidelines, including adherence to the legal requirements of the study country.

\section{References}
\bibliography{main_bib.bib}
\bibliographystyle{unsrt}

\end{document}